**Are the COVID19 restrictions really worth the cost? A comparison of estimated mortality in Australia from COVID19 and economic recession.**

*Neil W Bailey (1)*

*Daniel West (2)*

*\* Joint primary authorship*

(1) Epworth Centre for Innovation in Mental Health, Monash University
(2) Paxton Partners

**Abstract**

There has been considerable public debate about whether the economic impact of the current COVID19 restrictions are worth the costs. Although the potential impact of COVID19 has been modelled extensively, very few numbers have been presented in the discussions about potential economic impacts. For a good answer to the question "will the restrictions cause as much harm as COVID19?" credible evidence-based estimates are required, rather than simply rhetoric[1]. Here we provide some preliminary estimates to compare the impact of the current restrictions against the direct impact of the virus. Since most countries are currently taking an approach that reduces the number of COVID19 deaths, the estimates we provide for deaths from COVID19 are deliberately taken from the low end of the estimates of the infection fatality rate[2], while estimates for deaths from an economic recession are deliberately computed from double the high end of confidence interval for severe economic recessions. This ensures that an adequate challenge to the status quo of the current restrictions is provided. Our analysis shows that strict restrictions to eradicate the virus are likely to lead to at least eight times fewer total deaths than an immediate return to work scenario.

**Negative Impacts of the COVID19 restrictions**

Perhaps the most obvious component of the argument to lift restrictions is that a recession causes unemployment. Unemployment reduces mental health, resulting in increased suicide rates. This link has been studied at length in the academic literature[3].

Studies of the link between unemployment and suicide in 26 European countries over 40 years showed that increases in unemployment of more than 3% were associated with a 4.45% increase in suicides in those under 65[4]. Australian data also suggests that suicide rates are higher in those who are unemployed[5]. It is worth noting that despite our intuition and some evidence for the link between unemployment and suicide, the link is not confirmed across the scientific literature. At least one large scale analysis suggests the association between increased unemployment rates and suicide is not statistically significant in most models and samples, so the evidence for an increase in suicides from an increased unemployment rate is not conclusive[6].

As well as the effects on the economy, COVID19 restrictions cause general disruption to people's lives. One of the most significant risks is from the increase in loneliness[7]. Meta-analyses show that prolonged severe loneliness can increase all-cause mortality by 15-29%[8],[9]. It should be noted that this statistic comes from research examining severe and chronic loneliness, often associated with other factors such as depression, which also increases mortality rates. Very little research has looked at the effect of temporary increases in loneliness due to a situation like the current pandemic restrictions. Preliminary research from Poland suggests that in the first week of the COVID19 restrictions, only 6% reported spending the week alone (despite 80-90% compliance) and those who reported compliance with the restriction measures only reported a slight increase in loneliness compared to those who did not[10]. A recent review on the effects of quarantine suggests an increase in the number of people showing psychological distress by approximately 20% across multiple studies and measures, so we have used this higher value (even though quarantine due to contact with someone who is infected is likely to be more distressing than the population wide restrictions)[11]. We conservatively assumed severe restrictions continue for six months and used the estimate that 20% of Australians will experience loneliness during the restriction period. When multiplied by a 22% increase in the typical mortality rate per year (0.73%) for that 20% of the population (over six months) our calculations suggest up to 4,015 more deaths (with all these values probably reflecting very high estimates given the caveats listed above).

In addition to the increased mortality rate due to unemployment related suicides and loneliness, the current restrictions are likely to increase mortality from other factors that we do not currently have data on. Examples include increased domestic violence and under diagnosis of medical conditions as people delay visiting doctors. These effects might be matched in size by increased mortality through impacts on mental health, and result in people avoiding visiting their GPs for fear of infection. If hospitals are overwhelmed by COVID19 cases, deaths from injuries and illnesses other than COVID19 would also increase.

Lastly, it would also be reasonable to assume that a recession and high unemployment would increase the number of deaths from causes other than suicide. However, a number of large-scale studies suggest that "all-cause mortality" either did not change or even decreased during recessions with high unemployment[4],[12]. These analyses have indicated a drop in the overall mortality rate of 0.5% for every 1% increase in unemployment, and an increase in life expectancy even in European countries that were severely affected[13]. Studies found this to be the case even in Greece (which was amongst the worst affected)[14]. As such, for the purposes of our estimates, we assumed no change in the death rate due to causes other than suicide or loneliness.

**Scale of recession**

Current projections for the scale of this recession are grim. The pessimistic range of the estimates from the Grattan Institute has estimated the recession could result in as much as 15% unemployment[15]. The Grattan Institute's report notes that it took 7.6 years for employment to return to pre-recession levels after a recession in the 1990's, but suggests the recession resulting from COVID19 might not last as long. However, to provide a challenge to the status quo of restrictions, we have assumed it will take 10 years for unemployment levels to return to pre-COVID19 levels. To calculate the effect of this on the number of suicides, we calculated an estimate based on an increase by 16.48% in suicides (double the upper range of the 95% confidence interval for the 5% of the most severe recessions in Europe over a 40-year period[4]). As unemployment returns to the usual 5% over a 10-year period, we assume the suicide rate will decline proportionally across that time period. This pessimistic scenario would result in around 2,761 additional deaths from suicide over the ten-year period.

As far as we are aware, there are no reasonable projections for how much an economic recession will be mitigated by an immediate end to restrictions. However, the World Health Organisation and World Bank suggest approximately 60% of the economic costs from pandemics will be the result of both government and individual reactions to reduce the spread of the virus[16],[17]. If restrictions are lifted immediately, there will still be some individual reactions and government measures that will continue, which would not be otherwise be present in our typical economy. As such, for the sake of comparison, we have assumed that lifting restrictions immediately would mean the recession will result in half as many jobs lost (an increase from the current 5% unemployment rate to 10%, rather than the 15% we have assumed if restrictions continue). We have also assumed the recession will last only five years (rather than the ten years assumed if restrictions continue), and that this will result in only half the suicide rate.

Using these estimates, an increase to 10% unemployment decreasing over five years would result in 753 suicides (2,008 less than we have estimated from continued restrictions that leads to 15% unemployment).

These estimates are based on the hypothesis that we are choosing between the effects of restrictions to reduce the spread of the virus on the economy, and the lesser effects of the lifting of restrictions. As over 250 economists suggest, this is likely to be a false choice[18]. It is highly likely that the economy would be affected by the considerable number of deaths from COVID19 that would likely result from a release of restrictions. Evidence from the 1918 Spanish flu even suggests that cities that administered more severe restrictions during the pandemic had economies that bounced back faster after the pandemic than cities that did not administer severe restrictions[19]. This suggests that continued restrictions might benefit us both in terms of reducing the spread of COVID19 and reducing its impact on the economy. However, we will include the assumption that lifting restrictions would lead to faster economic recovery in our comparison, for the sake of providing an answer to the argument that places COVID19 restrictions in competition against the economy.

**COVID19 Modelling**

For a comparison between a recession from continued restrictions and the release of restrictions to protect the economy, we also provide estimates of the number of deaths due to COVID19. In doing this we have assumed that symptomatic individuals will be quarantined until they are no longer contagious, regardless of the approach taken. Modelling suggests removing all measures would result in 89% of Australians contract the virus[20]. This is not a reasonable scenario for comparison - even those advocating for an immediate return to work are seldom advocating for zero quarantine.

To estimate the death toll from different COVID19 approaches, the rate at which patients require critical care was projected to the Australian population by applying age-stratified findings from a recent study published in the Lancet[2]. This study indicated that there are much higher coronavirus infection rates in the population than is being revealed by current testing. As such, the rate of fatalities per infection is likely to be lower than the rate used in the Australian government modelling[20] and the UK modelling[21]. We used the lower rate of fatalities per infection to be prudent (given the substantial range in estimates of the fatality rate per infection throughout the current literature). Using the lower estimate also provides the strongest challenge to the status quo of the current restrictions. We provide estimates for three different scenarios. Immediate return to work (which has been proposed by some to reduce the damage to the economy), Herd Immunity (reduced restrictions to allow increased economic function but still allow coronavirus spread to slowly spread until enough people are infected that the virus is limited by herd immunity), and Eradication (restrictions are maintained until the virus is completely contained, then extensive track and trace to completely eliminate the virus).

**Immediate Return to Work**: Restriction measures are lifted, and people return to their typical work and leisure activities. Under this scenario, the Australian government has projected that 67.5% of people would contract the virus before herd immunity was achieved. Based on infection severity rates by age group[22], this would result in 317,000 people requiring ICU care. The average amount of time needed in ICU is ten days. The Government aims to increase the number of ICU beds to 7,000. If the infection were to spread over a 12-week period (this is even more conservative than the current modelling), more than 250,000 people who would require ICU care would not be able to receive it. Combining this figure with the number of deaths in those who do receive ICU care results in over 287,000 deaths. We assume that this approach will result in a recession lasting 5 years with 10% initial unemployment (and the associated 753 additional deaths from suicide).

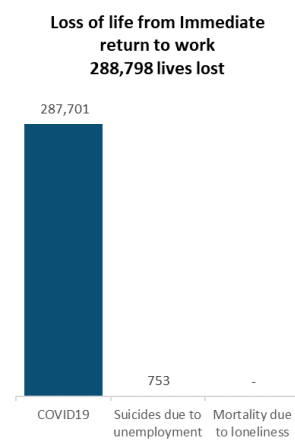

**Herd Immunity**: The current restriction measures of quarantine, isolation and social distancing remain in place (but are perhaps reduced) in order to "slow the spread" and eventually achieve herd immunity. Proponents of this scenario intend that the approach does not overwhelm the health system (so we assume that in this scenario everyone who needs ICU care receives it). To achieve herd immunity, approximately 60% of people need to contract the virus at some point (but perhaps higher based on other illnesses)[20],[23],[24]. Applying the estimates of the infection fatality rate by age group from the Lancet[2] study to the Australian population, there will be 282,000 Australians requiring ICU care and 141,000 deaths. To provide an adequate challenge to the status quo of continued restrictions, we have assumed that this approach still results in a deep recession of 10 years with 15% initial unemployment (and the associated 4,015 deaths from loneliness and 2,761 deaths from suicide).

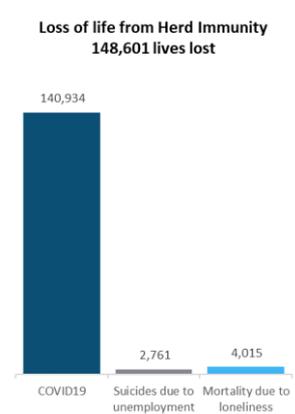

**Eradication**: The final scenario has been presented by the Grattan institute[25], who have stated that the end game must be total elimination of the virus. Current restriction measures would stay in place until all individuals who have been infected are no longer contagious. This is followed by strong border controls and diligent outbreak tracking to prevent further cases. Under this scenario (using the low infection fatality rates cited in the scenarios above), 11.6% of people would contract the virus and 27,000 would die[26]. As with the herd immunity strategy, we have assumed a deep recession over 10 years with 15% initial unemployment (and the associated 4,015 deaths from loneliness and 2,761 deaths from suicide). Worthy of note, early government modelling projected 11.6% infected as the best case, but progress to date in Australia seems well ahead of that. It is very possible that with continual prudent restrictions, the amount of deaths due to COVID19 will be well below 27,000.

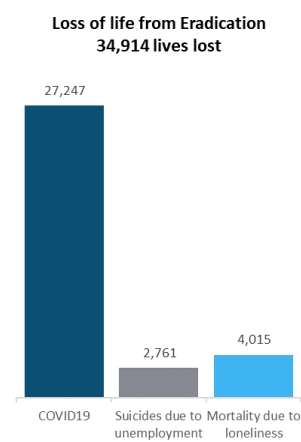

As noted above, the estimated infection fatality rate used in our calculations was conservative. If the infection fatality rate were based on the Australian and UK estimates, the estimates for COVID19 deaths would be 50% higher, resulting in 432,000 COVID19 related deaths for immediate return to work, 205,000 deaths for Herd immunity and 40,000 deaths for eradication.

**Further considerations**

The figures we have used for comparison are focused on mortality. Other important factors such as quality of life will vary between scenarios. These measures are more difficult to quantify and there is less available data, especially for the longer-term effects of COVID19. We have focused on deaths because they are generally recorded accurately and represent a critical consideration for health interventions like the current government restrictions.

Our estimates for the number of deaths from the Immediate Return to Work and from Herd Immunity scenarios also assume that individuals who have been infected are immune to reinfection. The WHO has stated that there is currently no evidence for this point[27], in which case the number of deaths from these two strategies could be considerably higher.

It is also worth noting that our estimates are extrapolated from data in developed countries. There is evidence that developing countries are more severely impacted by both pandemics[28] and recessions[29], so the conclusions from our estimates may not apply to all countries. There is also evidence that both pandemics and recessions disproportionally affect the vulnerable members of our society[30],[31],[32],[33]. In particular, young people are more likely to be affected by a recession[34]. Young people are also currently adhering to restrictions despite being at lower risk from COVID19. Government policies could be oriented towards ensuring younger persons are compensated for their involvement, given the asymmetry in risk to reward ratio compared to older individuals.

Lastly, the values provided in this article are forecasts based on previous data, and as such, while they are in line with the evidence, they may not eventuate as projected. Our estimates were biased towards a more favourable economy from immediate return to work, biased towards a higher number of potential deaths from a recession, and biased towards a lower number of deaths from COVID19. This approach was used to show the most favourable case for immediately returning to work, which provides the strongest challenge to the status quo of the restrictions.

**Conclusions**

In each of the modelled scenarios, there are three causes of death calculated: deaths from increased mortality caused by loneliness, deaths from an increased suicide rate associated with unemployment and deaths from COVID19 itself. It is worth noting that regardless of the strategy, deaths from COVID19 account for the largest proportion of deaths, far outweighing the number of deaths resulting from economic consequences.

As shown in the graph, each scenario resulted in a substantially different prediction of total number of deaths. Pursuing eradication may result in a harsh recession with many lives lost due to increased mortality associated with loneliness during the restriction period and due to unemployment over the next ten years. However, eradication also resulted in by far the lowest total number of deaths to COVID19 - 34,000.

Herd Immunity requires similar restrictions to eradication, as the spread of COVID19 must be slowed to a point where the health system can treat all those who need treatment. Even with ICU care available for all critical cases, 60% of the population would contract the virus and over 140,000 are estimated to die. Combined with the higher suicide rate the predicted total number of deaths is 148,000.

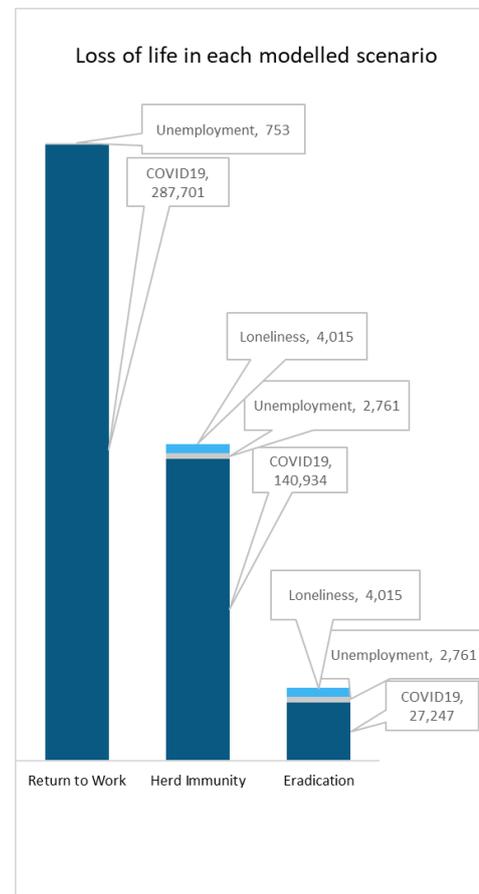

Despite assuming that immediately lifting restrictions would prevent all further deaths from loneliness and 72% of deaths from the increased suicide rate associated with high unemployment, the Return to Work immediately scenario is predicted to result in by far the highest overall number of deaths at 288,000. This is twice the number of deaths predicted for Herd Immunity and eight times as many as Eradication. It is worth noting that Sweden (which is applying looser restrictions than most countries and discussing herd immunity as the outcome), has already reported more deaths from COVID19 than the number of deaths we have estimated due to unemployment related suicide and loneliness (on a per capita basis). Perhaps this is why over 252 economists recently signed a letter recommending against loosening restrictions to benefit the economy[18].

While people are understandably concerned about their jobs, businesses, or investments, it seems highly likely that the cure (restrictions that prevent the virus from spreading) is not worse than the COVID19 disease. Continued restrictions to prevent the spread of coronavirus will lead to a far lower death toll, despite the potential negative effects from a retracted economy.

# Appendix A - COVID 19 Modelling outputs

## Immediate Return to Work

| Australian Hospital Capactiy | | Australian Government infection rate | | | Lancet article infection rate | | |
|---|---|---|---|---|---|---|---|
| Percentage contracting the virus | 67.5% | Bedday need | | 4,619,686 | Bedday need | | 3,171,009 |
| ICU beddays per patient | 10 | Gap | | 4,031,686 | Gap | | 2,583,009 |
| ICU capacity | 7,000 | Deaths for unmet need | | 403,169 | Deaths for unmet need | | 258,301 |
| period of spread (days) | 84 | ICU patients | | 58,800 | ICU patients | | 58,800 |
| Total beddays available | 588,000 | Deaths in ICU | | 29,400 | Deaths in ICU | | 29,400 |
| | | Total Deaths | | 432,569 | Total Deaths | | 287,701 |
| Age | Population | Age | Hospitalised | ICU required | | Hospitalised | ICU |
| 0-9 | 3,185,951 | 0-9 | 1,333 | 387 | | 239 | 69 |
| 10-19 | 3,058,048 | 10-19 | 1,280 | 372 | | 988 | 287 |
| 20-29 | 3,667,672 | 20-29 | 19,310 | 5,694 | | 5,189 | 1,530 |
| 30-39 | 3,672,564 | 30-39 | 71,890 | 21,071 | | 14,277 | 4,185 |
| 40-49 | 3,273,785 | 40-49 | 112,700 | 33,147 | | 24,193 | 7,116 |
| 50-59 | 3,079,754 | 50-59 | 205,805 | 60,286 | | 84,451 | 24,738 |
| 60-69 | 2,613,037 | 60-69 | 273,389 | 80,253 | | 231,930 | 68,083 |
| 70-79 | 1,792,241 | 70-79 | 433,095 | 127,025 | | 353,076 | 103,556 |
| 80+ | 1,021,255 | 80+ | 454,280 | 133,733 | | 365,297 | 107,538 |

## Herd Immunity

Percentage contracting the virus  60%

| | Australian Government infection rate | | | Lancet article infection rate | | |
|---|---|---|---|---|---|---|
| Age | Hospitalised | ICU | Deaths | Hospitalised | ICU | Deaths |
| 0-9 | 1,185 | 344 | 172 | 212 | 62 | 31 |
| 10-19 | 1,138 | 330 | 165 | 878 | 255 | 128 |
| 20-29 | 17,165 | 5,061 | 2,531 | 4,612 | 1,360 | 680 |
| 30-39 | 63,903 | 18,730 | 9,365 | 12,690 | 3,720 | 1,860 |
| 40-49 | 100,178 | 29,464 | 14,732 | 21,505 | 6,325 | 3,162 |
| 50-59 | 182,937 | 53,588 | 26,794 | 75,067 | 21,989 | 10,995 |
| 60-69 | 243,012 | 71,336 | 35,668 | 206,160 | 60,518 | 30,259 |
| 70-79 | 384,973 | 112,911 | 56,456 | 313,845 | 92,049 | 46,025 |
| 80+ | 403,804 | 118,874 | 59,437 | 324,709 | 95,589 | 47,795 |
| Total | | | 205,319 | | | 140,934 |

## Eradication

Percentage contracting the virus  11.6%

| | Australian Government infection rate | | | Lancet article infection rate | | |
|---|---|---|---|---|---|---|
| Age | Hospitalised | ICU | Deaths | Hospitalised | ICU | Deaths |
| 0-9 | 229 | 67 | 33 | 41 | 12 | 6 |
| 10-19 | 220 | 64 | 32 | 170 | 49 | 25 |
| 20-29 | 3,319 | 979 | 489 | 892 | 263 | 131 |
| 30-39 | 12,355 | 3,621 | 1,811 | 2,453 | 719 | 360 |
| 40-49 | 19,368 | 5,696 | 2,848 | 4,158 | 1,223 | 611 |
| 50-59 | 35,368 | 10,360 | 5,180 | 14,513 | 4,251 | 2,126 |
| 60-69 | 46,982 | 13,792 | 6,896 | 39,858 | 11,700 | 5,850 |
| 70-79 | 74,428 | 21,829 | 10,915 | 60,677 | 17,796 | 8,898 |
| 80+ | 78,069 | 22,982 | 11,491 | 62,777 | 18,481 | 9,240 |
| Total | | | 39,695 | | | 27,247 |

[1] Gong, B., Zhang, S., Yuan, L., & Chen, K. Z. (2020). A balance act: minimizing economic loss while controlling novel coronavirus pneumonia. *Journal of Chinese Governance*, 1-20.

[2] Verity, R., Okell, L. C., Dorigatti, I., Winskill, P., Whittaker, C., Imai, N., ... & Dighe, A. (2020). Estimates of the severity of coronavirus disease 2019: a model-based analysis. *The Lancet Infectious Diseases*.

[3] Oyesanya, M., Lopez-Morinigo, J., & Dutta, R. (2015). Systematic review of suicide in economic recession. *World journal of psychiatry*, *5*(2), 243.

[4] Stuckler, D., Basu, S., Suhrcke, M., Coutts, A., & McKee, M. (2009). The public health effect of economic crises and alternative policy responses in Europe: an empirical analysis. *The Lancet*, *374*(9686), 315-323.

[5] Milner, A., Morrell, S., & LaMontagne, A. D. (2014). Economically inactive, unemployed and employed suicides in Australia by age and sex over a 10-year period: what was the impact of the 2007 economic recession? *International journal of epidemiology*, *43*(5), 1500-1507.

[6] Tapia Granados, J. A., & Ionides, E. L. (2017). Population health and the economy: Mortality and the Great Recession in Europe. *Health economics*, *26*(12), e219-e235.

[7] Carrion, V. G., McCurdy, B. H., & Scozzafava, M. D. Increased Risk of Suicide Due to Economic and Social Impacts of Social Distancing Measures to Address the Covid-19 Pandemic: A Forecast.

[8] Rico-Uribe, L. A., Caballero, F. F., Martín-María, N., Cabello, M., Ayuso-Mateos, J. L., & Miret, M. (2018). Association of loneliness with all-cause mortality: A meta-analysis. *PloS one*, *13*(1).

[9] Holt-Lunstad, J., Smith, T. B., Baker, M., Harris, T., & Stephenson, D. (2015). Loneliness and social isolation as risk factors for mortality: a meta-analytic review. *Perspectives on psychological science*, *10*(2), 227-237.

[10] Okruszek, L., Aniszewska-Stańczuk, A., Piejka, A., Wiśniewska, M., & Żurek, K. (2020). Safe but lonely? Loneliness, mental health symptoms and COVID-19.

[11] Brooks, S. K., Webster, R. K., Smith, L. E., Woodland, L., Wessely, S., Greenberg, N., & Rubin, G. J. (2020). The psychological impact of quarantine and how to reduce it: rapid review of the evidence. *The Lancet*.

[12] Strumpf, E. C., Charters, T. J., Harper, S., & Nandi, A. (2017). Did the Great Recession affect mortality rates in the metropolitan United States? Effects on mortality by age, gender and cause of death. *Social Science & Medicine*, *189*, 11-16.

[13] Tapia Granados, J. A., & Ionides, E. L. (2017). Population health and the economy: Mortality and the Great Recession in Europe. *Health economics*, *26*(12), e219-e235.

[14] Laliotis, I., Ioannidis, J. P., & Stavropoulou, C. (2016). Total and cause-specific mortality before and after the onset of the Greek economic crisis: an interrupted time-series analysis. *The Lancet Public Health*, *1*(2), e56-e65.

[15] Coates, B., Cowgill, M., Chen, T., & Mackey, W. (2020). Shutdown: estimating the COVID-19 employment shock.

[16] Fan, V. Y., Jamison, D. T., & Summers, L. H. (2018). Pandemic risk: how large are the expected losses?. *Bulletin of the World Health Organization*, *96*(2), 129.

[17] Jonas, O. B. (2014). Pandemic Risk. World Development Report.

[18] https://theconversation.com/open-letter-from-265-australian-economists-dont-sacrifice-health-for-the-economy-136686

[19] Correia, Sergio and Luck, Stephan and Verner, Emil, Pandemics Depress the Economy, Public Health Interventions Do Not: Evidence from the 1918 Flu (March 30, 2020). Available at SSRN: https://ssrn.com/abstract=3561560 or http://dx.doi.org/10.2139/ssrn.3561560.

[20] Moss, R., Wood, J., Brown, D., Shearer, F., Black, A. J., Cheng, A., ... & McVernon, J. (2020). Modelling the impact of COVID-19 in Australia to inform transmission reducing measures and health system preparedness. *medRxiv*.

[21] Ferguson, N., Laydon, D., Nedjati Gilani, G., Imai, N., Ainslie, K., Baguelin, M., ... & Dighe, A. (2020). Report 9: Impact of non-pharmaceutical interventions (NPIs) to reduce COVID19 mortality and healthcare demand.

[22] Moss, R., Wood, J., Brown, D., Shearer, F., Black, A. J., Cheng, A., ... & McVernon, J. (2020). Modelling the impact of COVID-19 in Australia to inform transmission reducing measures and health system preparedness. *medRxiv*.

[23] Paulke-Korinek, M., Kundi, M., Rendi-Wagner, P., de Martin, A., Eder, G., Schmidle-Loss, B., ... & Kollaritsch, H. (2011). Herd immunity after two years of the universal mass vaccination program against rotavirus gastroenteritis in Austria. *Vaccine*, *29*(15), 2791-2796.

[24] Panagiotopoulos, T., Berger, A., & Valassi-Adam, E. (1999). Increase in congenital rubella occurrence after immunisation in Greece: retrospective survey and systematic review. How does herd immunity work?. *Bmj*, *319*(7223), 1462-1467.

[25] https://grattan.edu.au/news/australias-endgame-must-be-total-elimination-of-covid-19/

[26] Moss, R., Wood, J., Brown, D., Shearer, F., Black, A. J., Cheng, A., ... & McVernon, J. (2020). Modelling the impact of COVID-19 in Australia to inform transmission reducing measures and health system preparedness. *medRxiv*.

[27] World Health Organization. (2020). *"Immunity passports" in the context of COVID-19: scientific brief, 24 April 2020* (No. WHO/2019-nCoV/Sci_Brief/Immunity_passport/2020.1). World Health Organization.

[28] Oshitani, H., Kamigaki, T., & Suzuki, A. (2008). Major issues and challenges of influenza pandemic preparedness in developing countries. *Emerging infectious diseases*, *14*(6), 875.

[29] Ocampo, J. A., Griffith-Jones, S., Noman, A., Ortiz, A., Vallejo, J., & Tyson, J. (2012). The great recession and the developing world. *Development Cooperation in Times of Crisis*, 17-81.

[30] Wang, Y., Di, Y., Ye, J., & Wei, W. (2020). Study on the public psychological states and its related factors during the outbreak of coronavirus disease 2019 (COVID-19) in some regions of China. *Psychology, Health & Medicine*, 1-10.

[31] Liu, D., Ren, Y., Yan, F., Li, Y., Xu, X., Yu, X., ... & Yao, Y. (2020). Psychological Impact and Predisposing Factors of the Coronavirus Disease 2019 (COVID-19) Pandemic on General Public in China.

[32] Wang, C., Pan, R., Wan, X., Tan, Y., Xu, L., Ho, C. S., & Ho, R. C. (2020). Immediate psychological responses and associated factors during the initial stage of the 2019 coronavirus disease (COVID-19) epidemic among the general population in China. *International journal of environmental research and public health*, *17*(5), 1729.

[33] Holland, P., Burström, B., Whitehead, M., Diderichsen, F., Dahl, E., Barr, B., ... & Clayton, S. (2011). How do macro-level contexts and policies affect the employment chances of chronically ill and disabled people? Part I: The impact of recession and deindustrialization. *International Journal of Health Services*, *41*(3), 395-413.

[34] Nesvisky, M. (2012). The career effects of graduating in a recession.